\RequirePackage{arydshln}
\documentclass[aps,twocolumn,nofootinbib,superscriptaddress,preprintnumbers,pra,10pt,floatfix]{revtex4-1}

\usepackage[utf8]{inputenc}

\usepackage{float}
\usepackage{amsmath,amssymb}
\usepackage{dsfont}
\usepackage{hyperref}
\usepackage{graphicx}
\usepackage{enumitem}
\usepackage{mathtools}
\usepackage{slashed}
\usepackage{bbold}
\usepackage{multirow}
\usepackage{ytableau}
\usepackage{youngtab}
\usepackage{braket}
\usepackage{soul}
\usepackage{cancel}
\usepackage{xcolor}
\usepackage[normalem]{ulem}

\usepackage{lipsum}

\usepackage{colortbl} 
\definecolor{Gray}{gray}{0.95}
\definecolor{RGray}{gray}{0.90}
\definecolor{CGray}{gray}{0.92}
\definecolor{nicegreen}{rgb}{0.1,0.5,0.1}
\definecolor{nicepurple}{rgb}{0.9,0.,0.9}

\usepackage{arydshln}

\newcommand{\mc}{\mathcal}
\makeatletter
\g@addto@macro\bfseries{\boldmath}
\makeatother

\makeatletter
\renewcommand\paragraph{\@startsection{paragraph}{4}{\z@}%
                                    {3.25ex \@plus1ex \@minus.2ex}%
                                    {-1em}%
                                    {\normalfont\normalsize\bfseries}}
\makeatother

\allowdisplaybreaks
\begin{document}

\preprint{}
\preprint{}

\title{Probing the neutrino mass through semileptonic meson decays}

\author{Damir Be\v{c}irevi\'c}
\email{damir.becirevic@ijclab.in2p3.fr}
\affiliation{IJCLab, P\^ole Th\'eorie (Bat.~210), CNRS/IN2P3 et Universit\'e Paris-Saclay, 91405 Orsay, France}
\author{Claire Chevallier}
\email{claire.chevallier@ijclab.in2p3.fr}
\affiliation{IJCLab, P\^ole Th\'eorie (Bat.~210), CNRS/IN2P3 et Universit\'e Paris-Saclay, 91405 Orsay, France}
\author{Svjetlana Fajfer} 
\email{svjetlana.fajfer@ijs.si}
\affiliation{Department of Physics, University of Ljubljana, Jadranska 19, 1000 Ljubljana, Slovenia}
\affiliation{Jo\v zef Stefan Institute, Jamova 39, 1000  Ljubljana, Slovenia}
\author{Nejc Košnik} 
\email{nejc.kosnik@ijs.si}
\affiliation{Department of Physics, University of Ljubljana, Jadranska 19, 1000 Ljubljana, Slovenia}
\affiliation{Jo\v zef Stefan Institute, Jamova 39, 1000  Ljubljana, Slovenia}
\author{Lovre Pavičić} 
\email{lovre.pavicic@ijs.si}
\affiliation{Jo\v zef Stefan Institute, Jamova 39, 1000 Ljubljana, Slovenia}

\begin{abstract}
\vspace{5mm}
We argue that a detailed analysis of semileptonic decays can test the possibility of a massive neutrino. The key observable, related to the forward–backward asymmetry, is exactly zero for a massless neutrino but becomes non-zero if the neutral lepton is heavy and interacts with Standard Model fields via left-handed operators. For right-handed interactions, this quantity differs significantly from zero even for a massless right-handed neutrino. We demonstrate this explicitly using the example of a pseudoscalar meson decaying into another pseudoscalar meson. A similar discussion applies to decays into a vector meson, with an additional subtlety addressed in this work.
\vspace{3mm}
\end{abstract}

\maketitle

\allowdisplaybreaks

\section{Introduction}\label{sec:intro}
In recent years, a considerable amount of work has been devoted to investigating the possibility of probing massive neutrinos in low-energy experiments. In scenarios involving one or more sterile neutrinos, the inverse seesaw mechanism provides an elegant alternative for neutrino mass generation and opens the possibility of detecting a neutral lepton in the MeV--GeV energy window~\cite{Dasgupta:2021ies, Abdullahi:2022jlv, Bondarenko:2018ptm, Drewes:2013gca, Fernandez-Martinez:2023phj}.
In certain other Beyond the Standard Model (BSM) scenarios, new neutrino species are introduced to reconcile theoretical predictions with discrepancies observed between Standard Model (SM) calculations and experimental measurements~\cite{Felkl:2021uxi, Becirevic:2024iyi, Rosauro-Alcaraz:2024mvx, Bolton:2025fsq}. Even though these deviations are not yet significant at the multi-sigma level, scenarios in which the SM particle content is extended by an additional neutral lepton are intriguing; their underlying assumptions should be scrutinized through detailed comparisons between theory and experiment. In BSM models involving low-energy leptoquark states, such as $\overline{S}_1$ and/or $\widetilde R_2$, an additional neutral lepton (neutrino) may couple to the SM fields via left- or right-handed operators~\cite{Dorsner:2016wpm}. Light, right-handed, sub-GeV neutrinos are also plausible in left-right symmetric extensions of the SM~\cite{Abdullahi:2022jlv,Akhmedov:1995vm}.

In this paper, we argue that the presence of such neutrino species can be tested through the semileptonic decays of mesons or baryons. To make our examples transparent, we focus on the semileptonic decay of a pseudoscalar meson into another pseudoscalar meson. Similar decays into a vector meson offer a wider array of observables and, consequently, more opportunities to isolate the effects of a non-SM neutrino. Specifically, these observables can distinguish whether the dominant couplings are to a right-handed (RH) or a left-handed (LH) neutrino. A complementary approach to this possibility was recently explored in Ref.~\cite{Bernlochner:2024xiz}, which emphasized operators relevant to a massive RH neutrino~\cite{Robinson:2018gza}. In this work, we take a different route, as detailed in the following sections.

\section{Effective Theory and Observables}

The starting point of our discussion is the effective theory relevant to the semileptonic decays $d\to u \ell N$, where $u,d$ stand for a generic up- and down-type quark, $\ell$ for a lepton, and $N$ for a neutral lepton which in the SM is $N=\nu$.~\footnote{For the charm quark one actually considers $u\to d \ell N$. Since the effective Hamiltonian is hermitian both cases are captured by Eq.~\eqref{eq:hamilt}.} More generally, in addition to the SM neutrino, we allow $N$ to be also either $N_L$ (a neutral lepton coupling to the SM fields exclusively via LH interactions) or $N_R$ (coupling via RH interactions). We treat  $N_{L,R}$ as massive in the general case, and the massless limit will be addressed when appropriate.  The corresponding low-energy effective theory (LEFT) Hamiltonian reads:
\begin{equation}\label{eq:hamilt}
\mathcal{H}_{\text{eff}} = \frac{4 G_F V_{ud}}{\sqrt{2}} 
\sum_{\substack{Y = S,V,T\\ A,B=L,R}}  C_{AB}^{Y} \mathcal{O}_{AB}^{Y} + \mathrm{h.c.}\,,
\end{equation}
where the generic four-fermion operators are:
\begin{equation}
	\label{eq:Ops}
	\begin{split}
\mathcal{O}_{AB}^{V} & = (\bar{u} \gamma^\mu P_A d) \, (\bar{\ell} \gamma_\mu P_B N), \\
\mathcal{O}_{AB}^{S} &= (\bar{u} P_A d) \, (\bar{\ell} P_B N), \\
\mathcal{O}_{AB}^{T} &= \delta_{AB} \, (\bar{u} \sigma^{\mu\nu} P_A d) \, (\bar{\ell} \sigma_{\mu\nu} P_A N),
	\end{split}
\end{equation}
with the usual projection operators $P_{L/R} = (1\mp \gamma_5)/2$, and with $N$ being either the SM neutrino $\nu$, or an additional neutral lepton, $N_{L,R}$.
Clearly, the SM ($N=\nu$) is easily recovered by taking $C^{V}_{LL} = 1$ and by setting all the other Wilson coefficients to zero. The main goal of this study is to seek 
the observables for which a non-zero value of one or more of the above Wilson coefficients can be experimentally distinguished. 
To do so, we will restrict our attention here to the semileptonic decays of a pseudoscalar meson to another pseudoscalar or to a vector meson. A similar discussion of baryon decays is straightforward and will not be discussed here.

\section{Decay to a pseudoscalar meson}

\begin{figure}[t!]
\centering
\includegraphics[width=1.\linewidth]{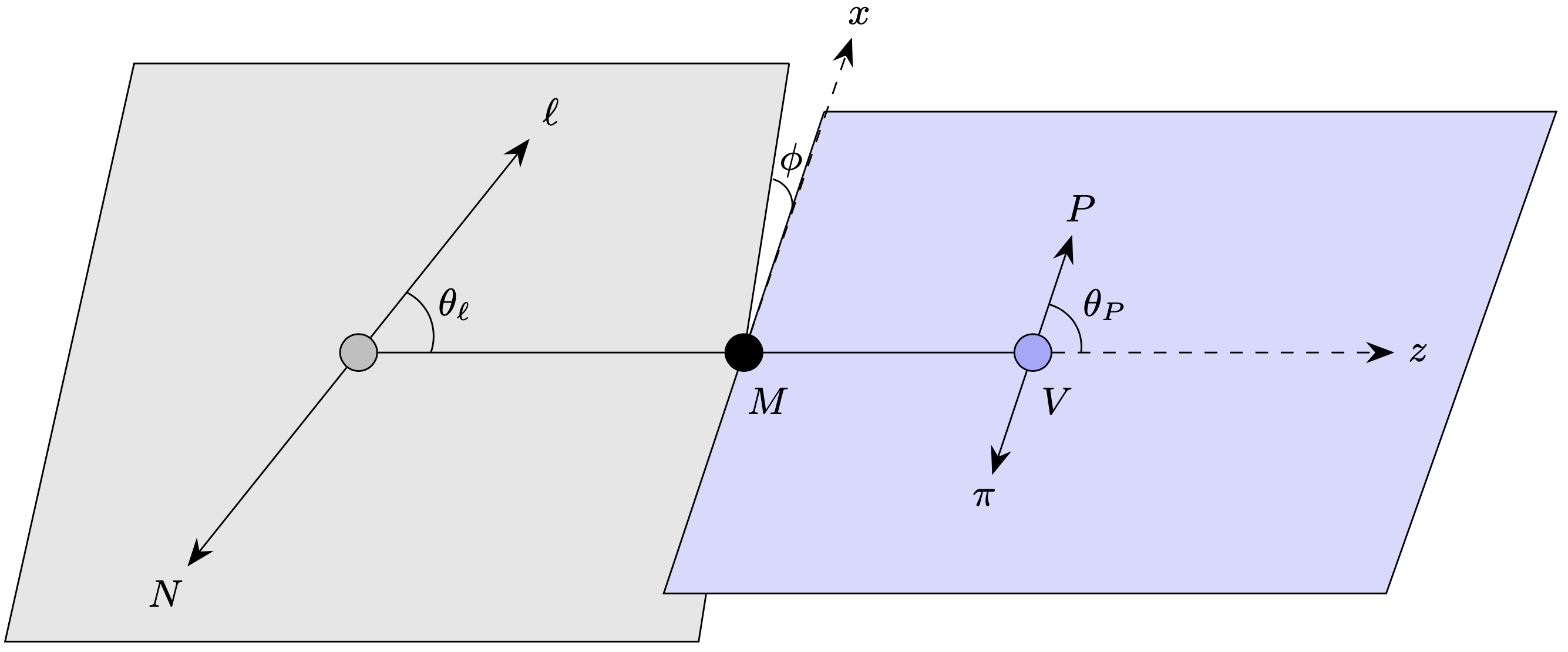}\\
\caption{\small \sl
Kinematics of the semileptonic decay of a pseudoscalar to a vector meson, $M\to V (\to P \pi) \ell N$. For a decay to a pseudoscalar meson, the angles $\phi$ and $\theta_P$ are not present.}
\label{fig:1} 
\end{figure}
A typical example of such a decay, that is being experimentally intensively studied in the past decade, is the case 
of $B\rightarrow D \ell \nu$. Several measurements of the ratios $R_{D^{(\ast )}}= 
\mathcal{B}(B \to D^{(\ast )} \tau\bar\nu)/\mathcal{B}(B \to D^{(\ast )} \mu\bar\nu)$ indicate a possibility of lepton flavor universality violation.  
The combined experimental results are $3\div 4\sigma$ larger than the values predicted in the SM~\cite{HeavyFlavorAveragingGroupHFLAV:2024ctg}. One way to explain this discrepancy 
is to evoke a possibility of a right handed neutrino ($N_R$) of mass $\mathcal{O}(\mathrm{MeV})$ or even $\mathcal{O}(\mathrm{GeV})$.

With the kinematics as indicated in Fig.~\ref{fig:1} and further specified in Appendix, the double differential decay width can be written as: 
\begin{equation}
	\label{eq:ang_coeff_P}
	\frac{d^2 \Gamma^s}{dq^2\ d\cos{\theta_\ell}} = a_s(q^2) + b_s(q^2) \cos{\theta_\ell} + c_s(q^2) \cos^2{\theta_\ell}\,,
\end{equation}
where $s\in \{+,-\}$ refers to the helicity of the lepton pair in the final state. We computed the above $q^2$-dependent functions and the results are collected in Appendix. 
It is then easy to obtain the full decay width as
 \begin{equation}
    \label{eq:main_text_gammaPPtot}
    \Gamma = \Gamma^+ + \Gamma^- = \int_{q_{\text{min}}^2}^{q_{\text{max}}^2} dq^2 \left[ 2 a(q^2) + \frac{2}{3} c(q^2) \right]\,,
\end{equation}
where $q_{\text{min}}^2 = (m_\ell + m_N)^2$ and $q_{\text{max}}^2 = (m_M - m_P)^2$, for a generic $M\rightarrow P \ell N$ semileptonic decay mode. 
Note also that $a(q^2) = a_-(q^2) + a_+(q^2)$, and similarly for the other two $q^2$-dependent functions, $b(q^2)$ and $c(q^2)$. 
Obviously, the coefficient function $b_s(q^2)$ disappears after the integration over $\theta_\ell$. The observable that directly isolates that function is the forward--backward asymmetry, 
\begin{align}\label{eq:AFB0}
\langle A_{FB}^s \rangle &= \frac{1}{\Gamma}  \int_{q_{\text{min}}^2}^{q_{\text{max}}^2}  dq^2 \left( \int_0^1 - \int_{-1}^{0} \right) d\cos\theta_\ell \, \frac{d^2\Gamma^s}{dq^2 d\cos\theta_\ell} \nonumber\\
&= \frac{1}{\Gamma}  \int_{q_{\text{min}}^2}^{q_{\text{max}}^2} dq^2\, b_{s}(q^2) \,.
\end{align}
By summing over the two helicities one recovers the full forward--backward asymmetry, $\langle A_{FB}  \rangle=\langle A_{FB}^- \rangle+\langle A_{FB}^+ \rangle$. 
In the above expressions we chose to integrate over the full available phase space, but one could also study the differential, $q^2$-dependent, forward--backward asymmetry which would 
in fact mean studying the shape of the function $b_s(q^2)$. An interesting feature, which was explicitly emphasized in Ref.~\cite{Becirevic:2020rzi}, is that in the SM 
\begin{equation}\label{b-SM}
b_-(q^2)^\mathrm{SM}=0\,,
\end{equation}
and therefore the forward--backward asymmetry is entirely due to $b_+(q^2)$.  This observation remains true even in the presence of other (BSM) operators 
in Eq.~\eqref{eq:Ops}, provided one works with a (massless) SM neutrino.  
For a non-zero $\langle A_{FB}^- \rangle$ the quark currents should either couple to (i) the LH massive neutral lepton, or (ii) to the  massive or \textbf{massless} RH neutral lepton.
In other words, by measuring $\langle A_{FB}^+ \rangle$ and $\langle A_{FB}^- \rangle$, and possibly their $q^2$-shapes, one can obtain a key constraint that can (in)validate 
the scenarios of physics BSM that involve additional neutral lepton species. We reiterate that the previous statement is valid for all the interaction operators given in Eqs.(\ref{eq:hamilt},\ref{eq:Ops}).

We have carefully studied all possible situations. In particular, if we restrict ourselves to the SM-like scenario of new physics, in which one only adds a massive neutral lepton (neutrino) without modifying the interaction Hamiltonian, one has $b_-(q^2)\neq 0$, and therefore measuring $\langle A_{FB}^- \rangle \neq 0$ would confirm such scenarios. Even more interesting is the case of a theory in which the SM particle content is extended by a massless RH neutrino. We find that in this situation as well $\langle A_{FB}^- \rangle \neq 0$. 

The presence of additional BSM operators makes the situation somewhat more complicated. Importantly, however, if only $C^{S}_{A,B}$ or $C^{T}_{A,B}$ are allowed to be non-zero, $\langle A_{FB} \rangle$ is not modified at all. A contribution to $\langle A_{FB} \rangle$ can be generated only if we have a SM-like scenario, or if $C^{S}_{A,B}$ or $C^{T}_{A,B}$ are not the only non-zero BSM Wilson coefficients. The resulting shift in $\langle A_{FB}^s \rangle$, however, turns out to be much smaller than in the case in which the Wilson coefficients $C^{V}_{AB}$ are chosen so as to be compatible with the measured branching fraction of a given semileptonic decay mode (within $2\sigma$). 

This remains true for semileptonic decays of kaons, $D$ mesons, and $B$ mesons. In the following, to make the discussion more specific, we consider scenarios in which the BSM contribution arises from couplings to the vector operator in Eq.~\eqref{eq:Ops}. We stress again that the expressions provided in the Appendix of this paper include contributions from all operators present in Eq.~\eqref{eq:Ops}.

We thus rewrite Eq.~\eqref{eq:hamilt} as  
\begin{align}\label{eq:hamilt2}
\mathcal{H}_{\text{eff}} =  \frac{4 G_F V_{ud}}{\sqrt{2}} \,  \sum_{B=L,R}\left(
C^V_{LB}\mathcal{O}_{LB}^V +C^V_{RB} \mathcal{O}^V_{RB} \right)\,,
\end{align}
where, again, the SM contribution ($N=\nu$) is recovered by setting $C_{LL}^V=1$. Otherwise the BSM vector operators comprise either the LH  or RH quark current and $B = (L,R)$, depending on the handedness of the neutral lepton field. We now discuss the case in which the interaction involves $N_L$, separately from the case involving $N_R$.

\subsection{LH neutral lepton}
We choose $B=L$ in Eq.~\eqref{eq:hamilt2} and compute the coefficients $b_{\pm}(q^2)$ entering Eq.~\eqref{eq:ang_coeff_P} for the semileptonic $M\to P\ell N_L$ decay. We find:
\begin{equation}
\label{eq:bL}
\begin{split}
    b_-(q^2)&=-\frac{ {\mathcal{N}}_{N_L}}{q^2}|C^V_{LL}+C^V_{RL}|^2 \\
    &\qquad \times  (m_M^2-m_P^2) \sqrt{\lambda(m_M^2,m_P^2,q^2)}\\
    &\qquad \times \text{Re}\left[f_+(q^2)f_0^*(q^2)\right] \left[K_{-+}(q^2,m_{N_L})\right]^2\\
    &=|C^V_{LL}+C^V_{RL}|^2 M(m_{N_L},q^2)\\
    &=M(0,q^2)+|C^V_{LL}+C^V_{RL}|^2 M(m_{N_L},q^2)\,,\\
    & \hfill \\
    b_+(q^2)&=\frac{m_M^2-m_P^2}{q^2} \sqrt{\lambda(m_M^2,m_P^2,q^2)}  \\
   &\qquad \times   \text{Re}\left[f_+(q^2)f_0^*(q^2)\right]  \left\{ {\mathcal{N}}_0 \left[K_{+-}(q^2,0)\right]^2 \right.  \\
  &\qquad \left. + {\mathcal{N}}_{N_L} |C^V_{LL}+C^V_{RL}|^2 
  \left[K_{+-}(q^2,m_{N_L})\right]^2\right\}\\ 
    &=P(0,q^2)+|C^V_{LL}+C^V_{RL}|^2P(m_{N_L},q^2)\,,
\end{split}
\end{equation}
where, in the last line of each equation we separate the SM contribution [$M(0,q^2)$ and $P(0,q^2)$], and we reiterate that $M(0,q^2)=0$, as already emphasized in Eq.~\eqref{b-SM}. We use the standard definition of the triangle function, $\lambda(m_a^2,m_b^2,q^2) =[q^2 - (m_a-m_b)^2 ][q^2 - (m_a+m_b)^2]$, that will be referred to as $\lambda_{ab}$ in the following. The normalization factor then reads,
\begin{equation}\label{eq:norm}
	{\mathcal{N}}_{N_L} = \frac{|V_{ud}|^2 G_F^2 }{128 \pi^3 m_M^3} \frac{\sqrt{\lambda_{MP}} \sqrt{\lambda_{\ell N_L}}}{4\, q^2  },
\end{equation} 
and ${\mathcal{N}}_{0}$ is trivially obtained from Eq.~\eqref{eq:norm} after setting $m_{N_L} =0$. 
In the notation of Ref.~\cite{Datta:2022czw}, which we adopt here, the functions $K_{\pm\pm}$ are given by,
\begin{equation}\label{eq:Kpm}
	K_{\pm\pm}(q^2,m_{N})\equiv \frac{(E_N+m_N\pm |p_\ell|)(E_\ell+m_\ell \pm |p_\ell |)}{\sqrt{(E_\ell +m_\ell)(E_N+m_N)}}\,,
\end{equation}
in the rest frame of the lepton pair. In the massless neutrino limit we have 
\begin{align}
&K_{++}(q^2,0) = 2\sqrt{q^2} \beta_\ell ,\quad K_{+-}(q^2,0) = 2m_\ell \beta_\ell\,, \cr 
&K_{-+}(q^2,0)= K_{--}(q^2,0) = 0\,, 
\end{align}
where $\beta_\ell = \sqrt{1-m_\ell^2/q^2}$. Note also that $f_{+,0}(q^2)$ in Eq.~\eqref{eq:bL} are the two semileptonic decay form factors 
which encode the nonperturbative QCD dynamics of the hadronic matrix element, 
\begin{align}
        \langle P(k)|\bar{u}\gamma^\mu d|M(p)\rangle = &\left[\left(p+k\right)^{\mu}-\frac{m_{P}^2-m_M^2}{q^2}q^\mu\right]f_+(q^2)\nonumber\\
        &+\frac{m_P^2-m_M^2}{q^2}q^{\mu}f_0(q^2)\,.
\end{align}

An important observation is that for our decays we have $q^2>m_\ell^2$, $m_M>m_P$, and since the form factors $f_{0,+}(q^2)$ are real and positive, the function $P(m_{N_L}, q^2)$ in Eq.~\eqref{eq:bL} is strictly positive, while the function $M(m_{N_L},q^2)$ is strictly negative. In other words, if the quark vector current couples to a massive LH neutral lepton field we have $\mathrm{sgn}[b_\pm(q^2)]=\pm 1$. When applied to the neutral current processes to two leptons in the final state, this feature will provide the change of sign of $A_{FB}(q^2)$: it is a combination of the strictly positive $A^+_{FB}(q^2)$ and the strictly negative $A^-_{FB}(q^2)$, and the two functions have different size and (slightly) different $q^2$-shapes.

\subsection{RH neutral lepton}

We now substitute $B=R$ in Eq.~\eqref{eq:hamilt2}, compute the functions $b_{\pm}(q^2)$ for $M\to P\ell N_R$, and find:
\begin{equation}
\label{eq:bR}
\begin{split}
    b_-(q^2)&= \frac{ {\mathcal{N}}_{N_R}}{q^2}|C^V_{LR}+C^V_{RR}|^2(m_M^2-m_P^2) \sqrt{\lambda_{MP}}\\
    &\qquad \times \text{Re}\left[f_+(q^2)f_0^*(q^2)\right] \left[K_{+-}(q^2,m_{N_R})\right]^2\\
    &=M(0,q^2)+ |C^V_{LR}+C^V_{RR}|^2 P(m_{N_R},q^2)\,,\\
    & \hfill \\
    b_+(q^2)&=\frac{m_M^2-m_P^2}{q^2} \sqrt{\lambda_{MP}} \, \text{Re}\left[f_+(q^2)f_0^*(q^2)\right]\times \\
  &\hspace*{-11mm} \left\{  {\mathcal{N}}_0 \left[K_{+-}(q^2,0)\right]^2 + {\mathcal{N}}_{N_R} |C^V_{LR}+C^V_{RR}|^2 
  \left[K_{-+}(q^2,m_{N_R})\right]^2\right\}\\ 
    &=P(0,q^2)+|C^V_{LR}+C^V_{RR}|^2 M(m_{N_R},q^2)\,.
\end{split}
\end{equation}
where $ {\mathcal{N}}_{N_R}$ is the same as in Eq.~\eqref{eq:norm} after replacing $N_L\to N_R$. In this situation $M(0,q^2)=0$, as explained before, cf. Eq.~\eqref{b-SM}. 
We see that for a massive $m_{N_R}$ the above expressions are simply flipped with respect to the expressions given in Eq.~\eqref{eq:bL}.  
By following the same line of arguments as in the previous subsection, and by keeping in mind that the two form factors are real and positive, we easily deduce 
that, this time, the function $b_-(q^2)$ is strictly positive. Therefore, 
a measurement of the sign of $\langle A_{FB}^-\rangle$ could help us distinguishing between the coupling to the LH from that to the RH neutral lepton. 
If the quark vector current couples to a massive LH neutrino then one has $\langle A_{FB}^-\rangle < 0$, while the coupling to the RH one results in $\langle A_{FB}^-\rangle > 0$.  
This information can be quite useful, provided $\langle A_{FB}^-\rangle$ can be measured. 
We stress once again that in the massless case, $b_-(q^2)=0$, for the LH neutrino, while it is non-zero and positive for the RH one. 

\section{Illustrative examples}

Let us now take two concrete examples. Various semileptonic decays allow us to probe different mass ranges. While the decay of $\overline K^0 \to \pi^+ \mu N$ 
can be helpful in checking for the presence of a heavy neutrino up to $m_N \lesssim 250$~MeV, the phase space for the similar $D^0 \to K^+ \mu N$ allows one to 
go up to $m_N \lesssim 1.26$~GeV. Finally, from the similar mode $\overline B^0 \to D^+  \mu N$ one can probe the mass of neutral lepton $m_N \lesssim 3.3$~GeV. 
An important feature of the $B$-decays is that there is enough phase space for $\tau$ in the final state. In the literature it is readily assumed that the decay $ B \to D  \tau N$ 
is responsible for the hints of lepton flavor universality violation, i.e. $R_D^\mathrm{exp} > R_D^\mathrm{SM}$, which is why the case of $B \to D  \tau N$ deserves a special attention~\cite{Becirevic:2024iyi,Azatov:2018kzb,Abada:2013aba,Cvetic:2017gkt}. 
Note also that the available phase space in $D^0 \to \pi^+  \mu N$ and $\overline B^0 \to D^+  \tau N$ are similar, both allowing to probe $m_N \lesssim 1.6$~GeV, and therefore 
if there is a breaking of lepton flavor universality due to the coupling of $\tau$ to the new massive neutrino species, such a feature can be checked by considering the two mentioned decay modes. 

To illustrate the discussion made in the previous section we will now use the experimental results from Ref.~\cite{ParticleDataGroup:2024cfk}, and the form factors from Ref.~\cite{FlavourLatticeAveragingGroupFLAG:2024oxs}. We employ the expressions given in Appendix, extract the bounds on the Wilson coefficients 
$C^V_{AB}$ from the experimental decay widths (to $2\sigma$) by switching on one Wilson coefficient at the time and then test the sensitivity of  
$\langle A_{FB}^\pm\rangle$ to the presence of $N_L$ or $N_R$. 

\begin{figure}[h!]
\centering
\hspace*{-3mm}\includegraphics[width=1.05\linewidth]{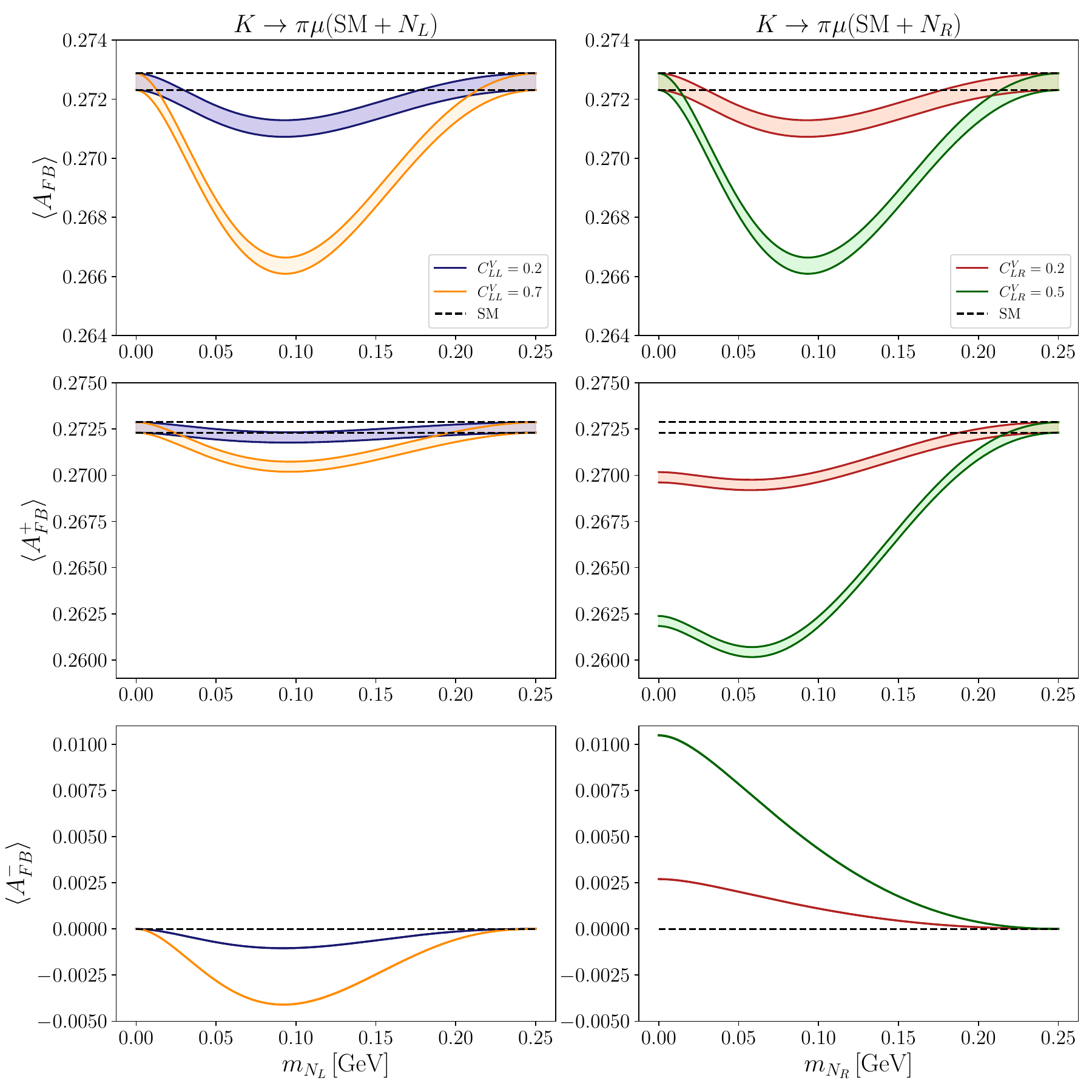}\\
\caption{\small \sl 
Forward--backward asymmetry of $\overline K^0 \to \pi^+ \mu \text{`inv'}$ integrated over the entire phase space, $\langle A_{FB}\rangle$, and its components corresponding to two helicities of the lepton pair,  $\langle A_{FB}^\pm\rangle$, all plotted as functions of $m_{N_L}$ ($m_{N_R}$). The dashed lines correspond to the SM values. }
\label{fig:2a} 
\end{figure}

In Figs.~\ref{fig:2a} we show that effect for the case of $\overline K^0 \to \pi^+ \mu \text{`inv'}$ where `inv' stands for the SM neutrino in addition to either $N_L$ or $N_R$ neutral lepton.~\footnote{An analogous plot is obtained in the case of $\overline D^0 \to K^+/\pi^+ \mu \text{`inv'}$.} 
Different curves correspond to different choices of the Wilson coefficients. 

We checked that the similar curves are obtained in the case of electron in the final state, but the whole effect is much smaller, and therefore it would be much more difficult to disentangle experimentally.

The effect shown in the plots can appear to be too small because we chose to normalize to the full decay rate. One can, instead, normalize to the polarized branching ratio and, define it bin-by-bin where in a bin of size $q^2\in [q_a^2,q_b^2]$,  one defines
\begin{equation} \label{eq:BES}
\left(  A_{FB}^\pm\right)^{[q^2_a,q^2_b]}_\mathrm{bin} = \frac{\displaystyle{ \int_{q_a^2}^{q_b^2} b_\pm (q^2) dq^2 }}{\displaystyle{\int_{q_a^2}^{q_b^2} {d\Gamma^\pm\over dq^2} dq^2 }}\,,
\end{equation}
just like it was recently done in Ref.~\cite{BESIII:2026ydr}. The effect we are trying to emphasize here would be more pronounced, even though the physics evidently remains the same. 
The normalization such as the one in Eq.~\eqref{eq:BES}, obviously assumes that a bin-by-bin analysis of these decays is experimentally feasible.

\begin{figure}[h!]
\centering
\hspace*{-6mm}\includegraphics[width=1.2\linewidth]{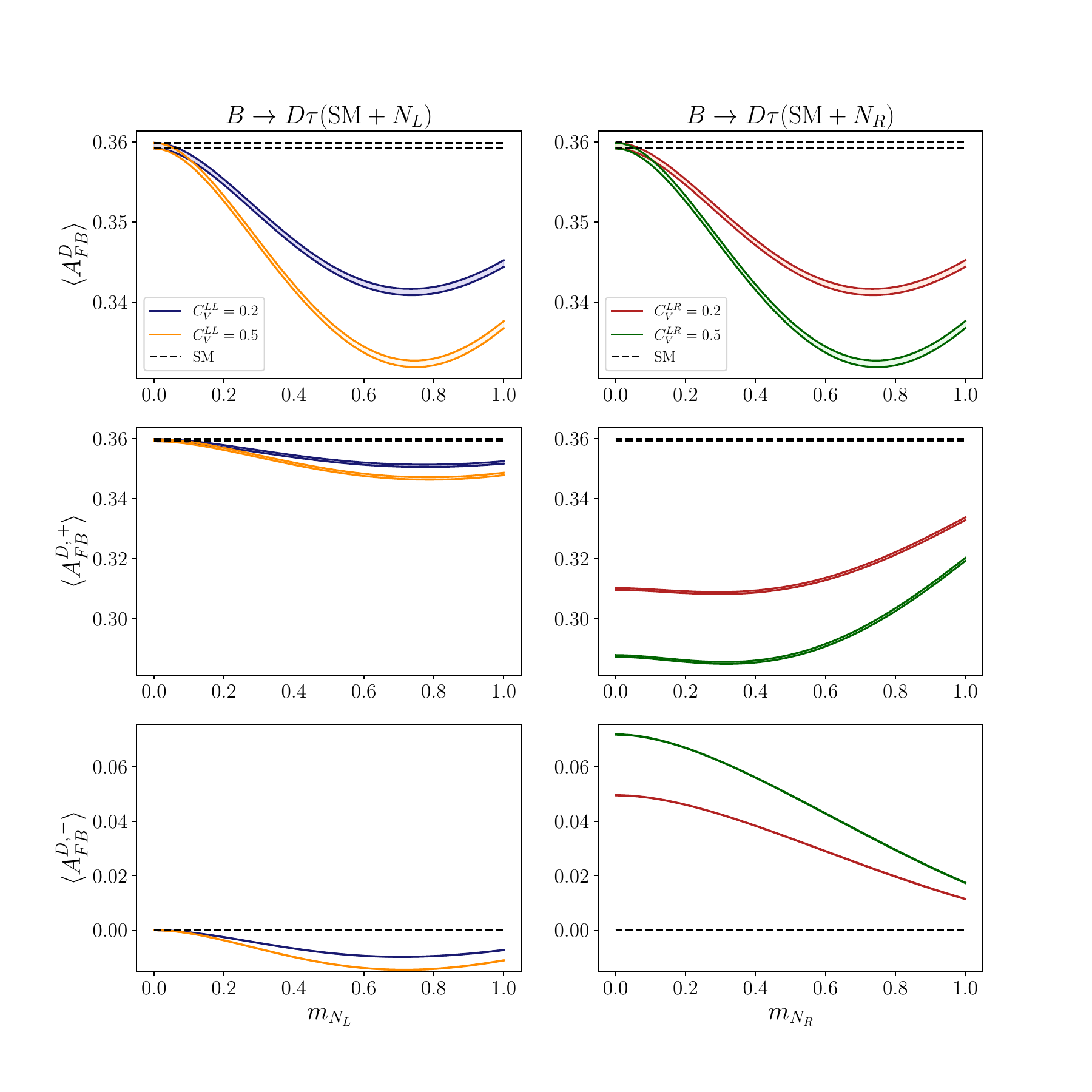}\\
\caption{\small \sl
Two components of the forward--backward asymmetry of $B \to D \tau \text{`inv'}$ corresponding to two helicities of the lepton pair. In the left (right) plots we show the dependence of the fully integrated $\langle A_{FB}^\pm\rangle$ as a function of the mass of $N_L$ ($N_R$). Plotted are the cases for two different values of $C_V^{LL}$ ($C_V^{LR}$). }
\label{fig:2b} 
\end{figure}

Another example, which is far more phenomenologically appealing and for which the effect of massive neutrino is more pronounced, is $B\to D\tau N$. We show the integrated $\langle A_{FB}^\pm\rangle$ in Fig.~\ref{fig:2b} for the case of SM being extended by $N_L$, separately from the case of $N_R$. Just like in the previous examples, we see that for the massless $N_L$ one recovers the SM case, but in the case of massless $N_R$ the effect is rather important and both $\langle A_{FB}^+\rangle$ and $\langle A_{FB}^-\rangle$ are significantly different from the SM case. 
Note also that we chose two values of the Wilson coefficient $C_V^{LL}$ ($C_V^{LR}$) that are consistent with the experimental value of $\mathcal{B}(B\to D\tau \text{`inv'})$. 
As for the experimental feasibility of such a study, we should mention that in Ref.~\cite{Belle:2017ilt} it was shown that $\Gamma^+$ and $\Gamma^-$ could be successfully separated. 
A new complexity is that in addition to that separation of events, an angular analysis should be done on each subset in order to expose the effect we are after in this paper.

\section{Decay to a vector meson}

\begin{figure}[h!]
\centering
\hspace*{-6mm}\includegraphics[width=1.192\linewidth]{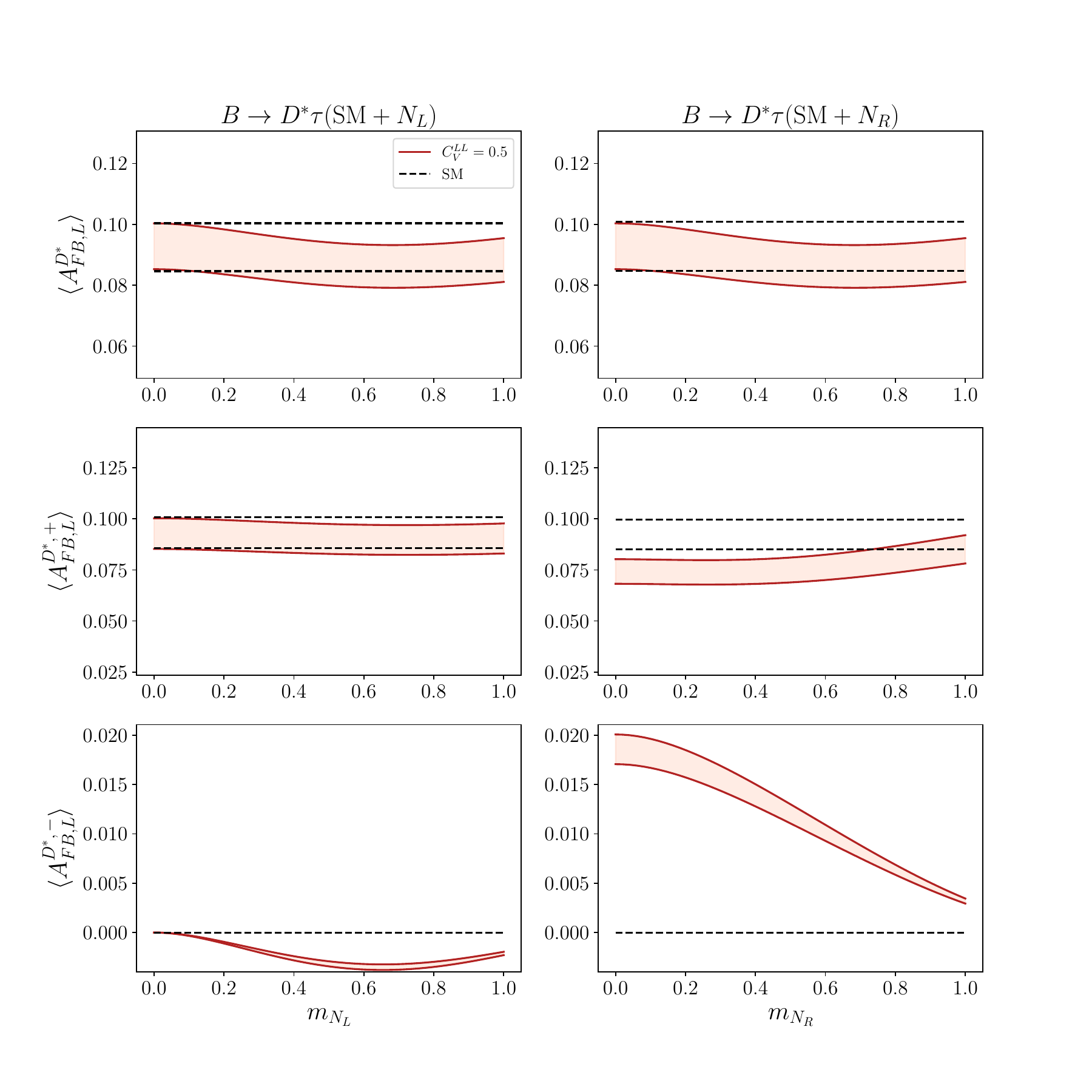}\\
\caption{\small \sl 
Two components of the forward--backward asymmetry of $B \to D^\ast \tau \text{`inv'}$ corresponding to the helicity of the outgoing leptons. In the left (right) plots we show the dependence of the fully integrated $\langle A_{FB,L}^\pm\rangle$ as a function of the mass of $N_L$ ($N_R$). Plotted are the cases for two different values of $C_V^{LL}$ ($C_V^{LR}$). }
\label{fig:3} 
\end{figure}

As it is well known, the decay to a vector meson provides many more observables due to a cascade decay to two pseudoscalar ones $V\to P\pi$. 
The angular analysis becomes more involved but since each coefficient (function of $q^2$) is independent, there would be as many observables as one can experimentally distinguish. 
Let us focus on the example of $B\to D^\ast (\to D\pi) \ell N_{L,R}$ and write the full angular distribution as~\cite{Bobeth:2021lya}:
\begin{equation}
	\label{eq:diff_w_Dst}
	\begin{aligned}
		\hspace*{-5mm}\frac{d^4 \Gamma^\pm}{dq^2 \, d \cos{\theta_\ell }\, d\cos{\theta_D}\, d\phi}  & =( I_{1s}^\pm + I_{2s}^\pm \cos{2\theta_\ell}+ I_{6s}^\pm \cos{\theta_\ell}) \sin^2{\theta_D} \\
		+& ( I_{1c}^\pm + I_{2c}^\pm \cos{2\theta_\ell}+ I_{6c}^\pm \cos{\theta_\ell}) \cos^2{\theta_D} \\
		+& (I_3^\pm \cos{2\phi} + I_9^\pm \sin{2\phi}) \sin^2{\theta_D} \sin^2{\theta_\ell}\\
		+& (I_4^\pm \cos{\phi} + I_8^\pm \sin{\phi}) \sin{2\theta_D} \sin{2\theta_\ell}\\
		+& (I_5^\pm \cos{\phi} + I_7^\pm \sin{\phi}) \sin{2\theta_D} \sin{\theta_\ell}\,,
	\end{aligned}
\end{equation}
where all of the coefficient functions $I^\pm_i = I^\pm_i (q^2)$ are computed in the LH or RH NP scenario and the expressions for those that are relevant to this paper are provided in Appendix. 
We again restrain our attention to the scenarios in which the SM is modified by $C^V_{LL}\neq 0$ or $C^V_{LR}\neq 0$, and check for the possibility to see the effects of 
LH or RH coupling to a massive or massless neutral lepton $N_L$ or $N_R$.  One can first integrate over $\phi\in[0,2\pi]$ and monitor the dependence on $\theta_D$, which for $\theta_D=0,\pi$ ($\theta_D=\pi/2$) would give rise to the contribution corresponding to the longitudinal (transverse) part of the decay rate. By identifying
\begin{equation}
	\begin{aligned}
		\hspace*{-0mm}\frac{d^4 \Gamma^\pm}{dq^2 \, d \cos{\theta_\ell }\, d\cos{\theta_D}}  & = \frac{3}{4} 
		\frac{d^3 \Gamma^\pm_T}{dq^2 \, d \cos{\theta_\ell }\, d\cos{\theta_D}} \sin^2 \theta_D  \\
		+&\frac{3}{2}\frac{d^3 \Gamma^\pm_L}{dq^2 \, d \cos{\theta_\ell }\, d\cos{\theta_D}} \cos^2 \theta_D ,	\end{aligned}
\end{equation}
one can easily rewrite each of the two pieces in terms of the functions $I^\pm_i$ from Eq.~\eqref{eq:diff_w_Dst}. The prefactors in the above expressions ensure that $\Gamma^\pm = \Gamma_L^\pm+\Gamma_T^\pm$. 
We again look for the forward--backward symmetry of $\Gamma_{L,T}$ and isolate the term multiplying $\cos\theta_\ell$, just like we did in Eq.~\eqref{eq:ang_coeff_P}. We find:
\begin{align}\label{eq:AFBLT}
b_\pm^L(q^2)= \frac{4\pi}{3}I_{6c}^\pm,\quad b_\pm^T(q^2)= \frac{8\pi}{3}I_{6s}^\pm.
\end{align}
After inspection we again observe that in the massless neutrino case, i.e. in the SM, 
\begin{align}
b_-^L(q^2)= 0\,, \quad b_+^T(q^2)= 0\,, 
\end{align}
and thus we are in the same situation as before, i.e. in the case of pseudoscalar meson in the final state. Analogously to Eq.~\eqref{eq:AFB0} we define:
\begin{align}\label{eq:AFBLT2}
\langle A_{FB,L(T)}^\pm \rangle &= \frac{1}{\Gamma_{L(T)}}  \int_{q_{\text{min}}^2}^{q_{\text{max}}^2} dq^2\, b_{\pm}^{L(T)}(q^2) \,.
\end{align}
with $q_{\text{max}}^2 =(m_\ell +m_{N_{L,R}})^2$ and $q_{\text{max}}^2=(m_B-m_{D^\ast})^2$. In Fig.~\ref{fig:3} we show $\langle A_{FB,L}^\pm \rangle$ as a function of $m_{N_{L(R)}}$. As discussed above, we indeed see that $\langle A_{FB,L}^- \rangle = 0$ in the SM and also in the case of an additional BSM neutral massless lepton that interacts via the left currents ($N_L$). As $m_{N_L}$ grows we 
see that the (longitudinal) forward--backward asymmetry becomes non-zero but strictly negative, $\langle A_{FB,L}^- \rangle < 0$. The effect in Fig.~\ref{fig:3} is small because we illustrated the case in which the normalization is made to a full decay rate. Had we chosen to work with a definition similar to Eq.~\eqref{eq:BES} the effect would have been more pronounced but still small for a realistic hope to be disentangled experimentally. 
Instead, and just like in the previous section, if the BSM neutral lepton is right handed, then $\langle A_{FB,L}^- \rangle > 0$, and the enhancement is the most significant in the massless case. The effect is nevertheless smaller than in the decay to a pseudoscalar meson in the final state, discussed in the previous Section.

\begin{figure}[h!]
\centering
\hspace*{-6mm}\includegraphics[width=1.192\linewidth]{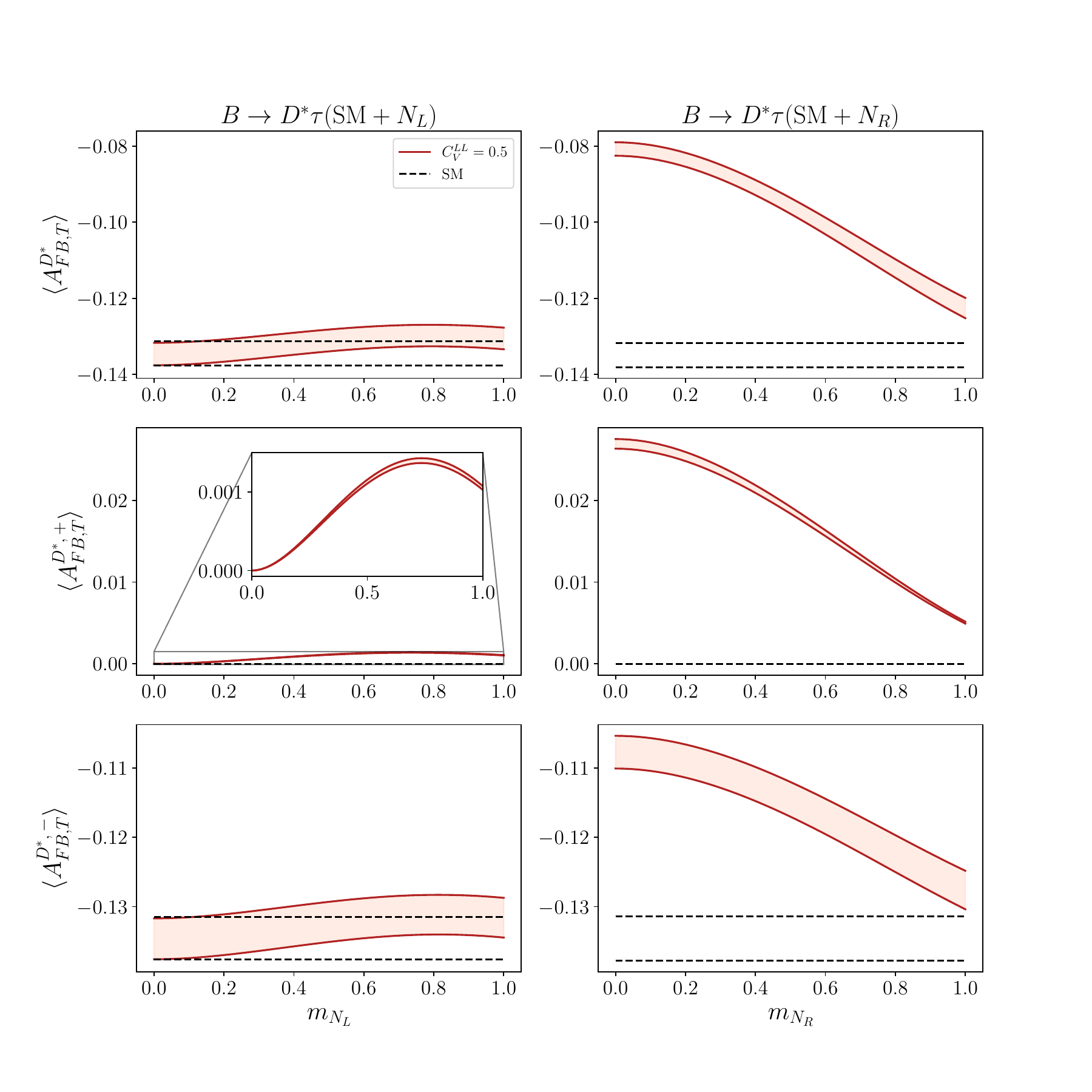}\\
\caption{\small \sl
Same as in Fig.~\ref{fig:3} but for $\langle A_{FB,T}^\pm\rangle$. }
\label{fig:4} 
\end{figure}

Finally, in Fig.~\ref{fig:4} we show the case of $\langle A_{FB,T} \rangle$, which in the SM gives $\langle A_{FB,T}^+ \rangle = 0$. The presence of a massive $N_L$ would provide a very small shift but for a massless or light $N_R$ the effect can be quite important. Note, however, that in this case changing the chirality of the quark current entails a change of the sign of $\langle A_{FB,T}^\pm \rangle$. We stress that, for the observables previously considered, changing the chirality of the quark current did not impact our results.
Experimentally, the longitudinal (transverse) fraction of the decay rate has been already measured~\cite{LHCb:2023ssl}. We understand that separating helicities could be challenging but hopefully eventually feasible as it would be beneficial for testing the scenarios with light massless neutrinos, whether they are participating in interactions via LH or RH currents.  

\subsection{More possibilities}
Since we computed the full angular distribution of the decay to a vector meson in which we separated the helicities of the outgoing leptons, we can check if some of the 
quantities are zero in the massless neutrino limit and become non-zero in the presence of a massive ${N_L}$ or either massive or massless ${N_R}$. 
We find three such quantities
\begin{align}
      S_c^{\pm} &= \frac{1}{\Gamma}\int_{q_{\text{min}}^2}^{q_{\text{max}}^2} dq^2\,  \left(I_{1c}^{\pm} + I_{2c}^{\pm}\right),\\
       S_s^{\pm} &= \frac{1}{\Gamma} \int_{q_{\text{min}}^2}^{q_{\text{max}}^2} dq^2\,  \left(I_{1s}^{\pm} + I_{2s}^{\pm}\right),\\
       D_s^{\pm} &= \frac{1}{\Gamma}\int_{q_{\text{min}}^2}^{q_{\text{max}}^2} dq^2\, \left(I_{1s}^{\pm} - 3 I_{2s}^{\pm}\right).
\end{align}
In the SM, $S_s^+=S_c^-=D_s^-=0$. The shifts in the case of massive $N_{L,R}$ is even smaller than in the cases discussed above. 

\section{Summary}\label{sec:summary}
In this paper we have shown that events in semileptonic decays with separated lepton polarization can provide a way to test the presence of a massive $\cal{O}(\mathrm{MeV}-\mathrm{GeV})$ neutrino participating in the interaction either through LH currents or RH ones. In particular, the forward--backward asymmetry of such separated events is useful because one of them is exactly zero in the SM ($m_\nu=0$) and becomes nonzero if the neutrino is massive (and left-handed). Moreover, the same forward--backward asymmetry becomes significantly different from zero even in the case of couplings to a massless RH neutrino. The sign of that asymmetry can also be used to determine whether the coupling to a massive neutral lepton occurs via a left- or right-handed current.

We illustrated how this works for the semileptonic decay of a pseudoscalar meson into another pseudoscalar meson. A similar discussion also applies to semileptonic decays into a vector meson, but in that case an additional separation of the events is required according to the polarization of the vector meson. Indeed, the decays into longitudinally and transversely polarized vector mesons separately exhibit the same feature as the one discussed above. In the examples we provided we restricted our attention to BSM vector operators, but we also presented the expressions for a generic scenario, i.e.\ within the low-energy effective theory in which, besides the vector operators, scalar and tensor operators are included.
\section*{Acknowledgments}
 S.F., N.K. and L.P. acknowledge financial support from the Slovenian Research Agency (research core funding No. P1-0035, N1-0321 and N1-0407).
This project has received support from the European Union’s Horizon 2020 research and innovation program under the Marie Sklodowska-Curie grant agreement No 860881-HIDDeN.

\section*{Appendix}
We provide the expressions for the semileptonic decays using the full set of operators given in LEFT~\eqref{eq:hamilt}. 
\subsection*{Decay to a pseudoscalar meson $M\to P \ell N$}
The kinematics in the rest frame of the lepton pair can be summarized as:
\begin{align}    
&p_M^\mu=(E_M,0,0,|\vec{p}_M|), ~ p_{P}^\mu=(E_{P},0,0,|\vec{p}_M|)\,,\cr
&p_N^{\mu}=(E_N, -|\vec{p}_\ell | \sin{\theta_\ell}, 0,-|\vec{p}_\ell| \cos{\theta_\ell})\,,\cr
		&p_\ell^{\mu}=(E_\ell, |\vec{p}_\ell| \sin{\theta_\ell}, 0,|\vec{p}_\ell| \cos{\theta_\ell})\,,
\end{align}
with $q^\mu=(p_N + p_\ell)^\mu = (\sqrt{q^2},0,0,0)$, $|\vec{p}_M|=\sqrt{\lambda_{MP}}/(2 \sqrt{q^2})$, $|\vec{p}_\ell|=\sqrt{\lambda_{N\ell}}/(2 \sqrt{q^2})$.
After computing the hadronic and leptonic amplitudes and putting them together, we write the resulting double differential decay rate in the form given in Eq.~\eqref{eq:ang_coeff_P}. 
To that end it is also useful to define the helicity amplitudes:
\begin{align}
    H_{V,0}(q^2) &= \frac{\sqrt{\lambda_{MP}}}{\sqrt{q^2}} f_+(q^2), \cr
    H_{V,t}(q^2) &= \frac{m_M^2 - m_P^2}{\sqrt{q^2}} f_0(q^2), \cr
    H_S(q^2) & =\frac{m_M^2 - m_P^2}{m_{q_d} - m_{q_u}} f_0(q^2), \cr
    H_T(q^2) &= -\frac{\sqrt{\lambda_{MP}}}{m_M + m_P} f_T(q^2),
\end{align}
where in the last amplitude the form factor $f_T(q^2)$ parametrizes the hadronic matrix element of the tensor density:
\begin{align}
    \label{eq:FT}
        \langle P(k)|\bar{u}\sigma^{\mu\nu}d|M(p)\rangle  = &-i\left(p^\mu k^\nu-p^\nu k^\mu\right)\frac{2\, f_T(q^2)}{m_M+m_P}\,.
\end{align}
We also need the helicity projector:
\begin{equation}
    \label{eq:hel_proj}
    P_\lambda = \frac{1-\lambda\, \slashed{s}\gamma_5}{2}\,,
\end{equation}
where $\lambda$ is positive or negative helicity and $s_\mu$ are chosen as 
\begin{equation}
\begin{aligned}
    s_\mu^{\ell} &= \frac{1}{m_\ell}(|\vec{p}_\ell|, E_\ell \frac{\vec{p}_\ell}{|\vec{p}_\ell|})\,\cr
    s_\mu^{N} &= \frac{1}{m_N}(|\vec{p}_\ell|, - E_N \frac{\vec{p}_\ell}{|\vec{p}_\ell|}).
\end{aligned}
\end{equation}
where we again used that $\vec{p}_\ell = - \vec{p}_N$.

After some algebra one gets the following expressions for the angular coefficients in Eq.~\eqref{eq:ang_coeff_P}:
	\begin{align}
	\label{eq:a pseudo}
		&a_- = 
		  \frac{\mathcal{N}_{N_L}}{2}\times\cr
		  &\Big[ \left| \left(C^S_{LL} + C^S_{RL} \right) H_S K_{--}^{N_L} + \left(C^V_{LL} + C^V_{RL} \right) H_{V,t} K_{-+}^{N_L} \right|^2 \cr
		& + \left| -4 C^T_{LL} H_T K_{+-}^{N_L} + \left( C^V_{LL} + C^V_{RL} \right) H_{V,0} K_{++}^{N_L} \right|^2\Big]\cr
		 &\qquad + \frac{\mathcal{N}_{N_R}}{2}\times\cr
		&\Big[
		 \left| \left(C^V_{LR} + C^V_{RR} \right) H_{V,0} K_{--}^{N_R} - 
		 4 C^T_{RR} H_T K_{-+}^{N_R} \right|^2 \cr 
		 & + \left| \left(C^V_{LR} + C^V_{RR} \right) H_{V,t} K_{+-}^{N_R} + 
		 \left(C^S_{LR} + C^S_{RR} \right) H_S K_{++}^{N_R} \right|^2
		 \Big]\,,\cr
       \end{align}
where the coefficient function $\mathcal{N}_{N}$ is given in Eq.~\eqref{eq:norm}, and $K_{\pm\pm}^N\equiv K_{\pm\pm}(q^2,m_{N})$ have been defined in Eq.~\eqref{eq:Kpm}.  
Note that the first term, corresponds to $N_L$, and the second to $N_R$, so that using the above expression for checking the response of the function $a_- \equiv a_-(q^2)$ to the presence of $N_L$ 
($N_R$) implies setting $\mathcal{N}_{N_R}$ ($\mathcal{N}_{N_L}$) to zero. 
Similarly, for the other two functions we have:	\begin{align}
	\label{eq:b pseudo}
		&b_- = \mathcal{N}_{N_L} 
		\times\cr 
		& \mathrm{Re} \Big\{ \Big[ (C^{S}_{LL} + C^{S}_{RL}) \, H_S \, K_{--}^{N_L} 
		+ ( C^{V}_{LL} + C^{V}_{RL}) \, H_{V,t} \, K_{-+}^{N_L}  \Big] \cr
		&\qquad  \Big[ 4 \, C^{T}_{LL} \, H_{T} \, K_{--}^{N_L} 
		-   ( C^{V}_{LL} + C^{V}_{RL})\, H_{V,0} \, K_{-+}^{N_L}\Big]\Big\} \cr
		&\qquad+\mathcal{N}_{N_R} \times\cr 
		 & \mathrm{Re} \Big\{\Big[ 
		 (C^{V}_{LR} + C^{V}_{RR}) \, H_{V,t} \, K_{+-}^{N_R} 
		+ (C^{S}_{LR} + C^{S}_{RR}) \, H_{S} \, K_{++}^{N_R}\Big] \cr
		&\Big[  (C^{V}_{LR} + C^{V}_{RR})H_{V,0} \, K_{+-}^{N_R} 
		- 4 \, C^{T}_{RR} \, H_{T} \, K_{++}^{N_R} \Big] \Big\}\,,\cr
        \end{align}

	\begin{align}
	\label{eq:c pseudo}
		&c_- = 
		  \frac{\mathcal{N}_{N_L}}{2}\times\cr
		  &\Big[ \left|4 C^T_{LL} H_T K_{--}^{N_L}  
		- ( C^V_{LL} + C^V_{RL}) H_{V,0} 
		K_{-+}^{N_L} \right|^2 \cr
		& -\left| 4 C^T_{LL} H_T  K_{+-}^{N_L} 
		-( C^V_{LL} + C^V_{RL}) H_{V,0} 
		 K_{++}^{N_L} \right|^2\Big]\cr
		 &\qquad + \frac{\mathcal{N}_{N_R}}{2}\times\cr
		&\Big[
		 \left| (C^V_{LR} + C^V_{RR}) H_{V,0} K_{--}^{N_R}  
		- 4 C^T_{RR} H_T K_{-+}^{N_R}  \right|^2  \cr 
		 & -
		\left| (C^V_{LR} + C^V_{RR}) H_{V,0}   K_{+-}^{N_R}
		- 4 C^T_{RR} H_T  K_{++}^{N_R} \right|^2
		 \Big]\,.\cr
       \end{align}
Similarly, for the helicity $s=+1$ we have:
	\begin{align}
	\label{eq:a+ pseudo}
		&a_+ = 
		  \frac{\mathcal{N}_{N_L}}{2}\times\cr
		  &\Big[ \left| ( C^V_{LL} + C^V_{RL}) H_{V,0} K_{--}^{N_L} 
		- 4 C^T_{LL} H_T K_{-+}^{N_L} \right|^2 \cr
		& + \left| ( C^V_{LL} + C^V_{RL}) H_{V,t} K_{+-}^{N_L} 
		+ (C^S_{LL} + C^S_{RL}) H_S K_{++}^{N_L} \right|^2 \Big]\cr
		 &\qquad + \frac{\mathcal{N}_{N_R}}{2}\times\cr
		&\Big[
		 \left| (C^V_{LR} + C^V_{RR}) H_{V,0} K_{++}^{N_R} 
		- 4 C^T_{RR} H_T K_{+-}^{N_R} \right|^2 \cr 
		 & + \left| (C^V_{LR} + C^V_{RR}) H_{V,t} K_{-+}^{N_R} 
		+ (C^S_{LR} + C^S_{RR}) H_S K_{--}^{N_R} \right|^2
		 \Big]\,,\cr
       \end{align}

	\begin{align}
	\label{eq:b+ pseudo}
		&b_+ = \mathcal{N}_{N_L} 
		\times\cr 
		& \mathrm{Re} \Big\{ \Big[  ( C^V_{LL} + C^V_{RL}) H_{V,t} K_{+-}^{N_L} 
		+ (C^S_{LL} + C^S_{RL}) H_S K_{++}^{N_L}\Big] \cr
		&\qquad  \Big[ ( C^V_{LL} + C^V_{RL}) H_{V,0} K_{+-}^{N_L} 
		- 4 C^T_{LL} H_T K_{++}^{N_L} \Big]\Big\} \cr
		&\qquad+\mathcal{N}_{N_R} \times\cr 
		 & \mathrm{Re} \Big\{\Big[ 
		 (C^V_{LR} + C^V_{RR}) H_{V,t} K_{-+}^{N_R} 
		+ (C^S_{LR} + C^S_{RR}) H_S K_{--}^{N_R} \Big] \cr
		&\Big[  -(C^V_{LR} + C^V_{RR}) H_{V,0} K_{-+}^{N_R} 
		+ 4 C^T_{RR} H_T  K_{--}^{N_R}  \Big] \Big\}\,,\cr
        \end{align}

	\begin{align}
	\label{eq:c+ pseudo}
		&c_+ = 
		  \frac{\mathcal{N}_{N_L}}{2}\times\cr
		  &\Big[ \left|- ( C^V_{LL} + C^V_{RL}) H_{V,0} 
		 K_{+-}^{N_L}  + 4 C^T_{LL} H_T 
		 K_{++}^{N_L}\right|^2 \cr
		& -\left|
		- ( C^V_{LL} + C^V_{RL})^2 H_{V,0} 
		K_{--}^{N_L}  + 4 C^T_{LL} H_T 
		 K_{-+}^{N_L}\right|^2\Big]\cr
		 &\qquad + \frac{\mathcal{N}_{N_R}}{2}\times\cr
		&\Big[
		 \left| -(C^V_{LR} + C^V_{RR}) H_{V,0}  K_{-+}^{N_R} 
		+ 4 C^T_{RR} H_T K_{--}^{N_R}  \right|^2  \cr 
		 & - \left| -(C^V_{LR} + C^V_{RR}) H_{V,0} K_{++}^{N_R}  
		+ 4 C^T_{RR} H_T K_{+-}^{N_R}  \right|^2
		 \Big]\,.\cr
       \end{align}
Note that in the text we focused on the scenario with the NP Wilson coefficients $C^V_{AB}\neq 0$ while all the other coefficients are left to zero. Obviously any other NP scenario can be easily 
implemented by using the general expressions given above.  Lastly, the SM expressions are obviously recovered by setting $m_{N_L}=0$, $\mathcal{N}_{N_R}=0$, $C^V_{LL}=1$ and all the other WIlson coefficients set to zero. 
\subsection*{Decay to a vector meson $M\to V(\to P\pi) \ell N$}
The kinematic of this process is very similar to the one discussed in the pseudoscalar case, except for the geometry that now involves two additional angles since the vector meson further decays $V\to D\pi$, as sketched in Fig.~\ref{fig:1}. 
We now have 
\begin{align}    
&p_M^\mu=(E_M,0,0,|\vec{p}_M|), ~ p_{V}^\mu=(E_{V},0,0,|\vec{p}_M|)\,,\cr
&p_N^{\mu}=(E_N,-|\vec{p}_\ell |\sin{\theta_\ell}\cos{\phi},-|\vec{p}_\ell |\sin{\theta_\ell}\sin{\phi},-|\vec{p}_\ell |\cos{\theta_\ell})\,,\cr
		&p_\ell^{\mu}=(E_\ell,|\vec{p}_\ell |\sin{\theta_\ell}\cos{\phi},|\vec{p}_\ell |\sin{\theta_\ell}\sin{\phi},|\vec{p}_\ell |\cos{\theta_\ell})\,,
\end{align}
and the decay amplitude can be written as 
\begin{align}
\mathcal{M}(M\rightarrow P \pi \ell N) = \mathcal{M}(M\rightarrow V  \ell N) \mathcal{M}( V \rightarrow P \pi) BW(m_{P\pi})\,
\end{align}
where the strong decay is described by  
\begin{align}
\mathcal{M} (V\rightarrow P \pi) &= g_{VP\pi} (p_P \cdot  \epsilon ),
\end{align}
with $\epsilon_\mu$ being the polarization vector of the vector meson and  $g_{VP\pi}$ is the strong interaction coupling that can be either computed in lattice QCD or extracted from experiment 
with the help of
\begin{align}
	\Gamma(V\rightarrow P\pi) = C_\pi \frac{\lambda_{VP\pi}^{3/2}}{192 \pi m_{V}^5} |g_{VP\pi}|^2 
\end{align}
where $C_{\pi^\pm}=1$, and $C_{\pi^0}=1/2$ due to isospin. In the rest frame of $V$, $p_P^\mu = (E_P,|p_P| \sin{\theta_P} ,0,|p_P| \cos{\theta_P})$, so that the amplitude is decomposed according to the polarization of $V$, to $\mathcal{M}_{P\pi}^0 = g_{VP\pi} |p_P| \cos{\theta_P}$ and $\mathcal{M}_{P\pi}^\pm = \pm (1/\sqrt{2}) g_{VP\pi} |p_P| \sin{\theta_P}$.~\footnote{More specifically, for the $D^\ast$ polarization vector $\epsilon_\lambda$ we chose: $\epsilon_0=(0,0,0,-1)$, $\epsilon_\pm=\mp (0,1,\pm i,0)/\sqrt{2}$. } 
In the narrow resonance approximation limit, $\Gamma_V\ll m_V$, 
\begin{equation}
	\label{eq:BW_approx}
	\begin{aligned}
		|BW(m_{P\pi})|^2 &= \frac{1}{(m_{P\pi}^2 - m_{V}^2)^2 - m_{V}^2 \Gamma_{V}^2 }\\
	&	\approx \frac{\pi}{m_{V} \Gamma_{V}} \delta(m_{P\pi}^2 - m_{V}^2)\,,
	\end{aligned}
\end{equation}
and thus it factorizes in the decay rate, and we can therefore focus on the decay to the vector meson alone, which is conveniently expressed as in Eq.~\eqref{eq:diff_w_Dst}. We computed the coefficient functions $I_i^\pm \equiv 
I_i^\pm(q^2)$ by using the LEFT~\eqref{eq:hamilt} and the results will be expressed in terms of the helicity amplitudes: 
	\begin{align}
	H^V_\pm =& (m_M+m_V) A_1(q^2) \mp \frac{\sqrt{\lambda_{MV}}}{m_M+m_V} V(q^2)\,,\cr
	H^V_0 =& -\frac{1}{2\sqrt{q2}} \Big[ (m_M+m_V)(m_M^2-m_V^2-q^2) A_1(q^2) - \cr 
	&\qquad \frac{\sqrt{\lambda_{MV}}}{m_M+m_V} A_2(q^2)\Big],\cr
	H^V_t =& - \frac{\sqrt{\lambda_{MV}}}{\sqrt{q^2}} A_0(q^2)\,,\cr
	H^S =& - \frac{\sqrt{\lambda_{MV}}}{m_u+m_d} A_0(q^2)\,,\cr
	H^T_0 =& -\frac{1}{2m_V} \Big[ (m_M^2+3 m_V-q^2) T_2(q^2) - \cr 
	&\qquad \frac{\sqrt{\lambda_{MV}}}{m_M^2-m_V^2} T_3(q^2)\Big],\cr
	H^T_\pm =& \frac{1}{\sqrt{q^2}} \Big[ \sqrt{\lambda_{MV}} T_1(q^2) - (m_M^2 - m_V^2) T_2(q^2)\Big]\,,\cr
	\end{align}
where we used the standard decomposition of the hadronic matrix elements in terms of form factors, namely,
\begin{align}
&\langle V(k, \lambda) | \bar u \gamma^\mu d | M(p) \rangle = -i \epsilon^{\mu \nu \rho \sigma} \epsilon_{\nu}^{(\lambda) \,\ast}  p_{\rho} k_{\sigma} \frac{2 V(q^2)}{m_M + m_V}, \cr
&\langle V(k, \lambda) | \bar u \gamma^\mu\gamma_5 d  | M(p) \rangle = q^\mu (\epsilon^*_{(\lambda)} \cdot q) \frac{2 m_V}{q^2} A_0(q^2)  \cr 
&\qquad + (m_M + m_V)  \left( \epsilon^{\mu *}_{(\lambda)} - q^\mu \frac{(\epsilon^*_{(\lambda_V)} \cdot q)}{q^2} \right) A_1(q^2) \cr
&\qquad - \frac{\epsilon^*_{(\lambda)} \cdot q}{m_M + m_V} A_2(q^2) \left( (p + k)^\mu - q^\mu \frac{m_M^2 - m_V^2}{q^2} \right), \cr
&\langle V(k, \lambda) |  \bar u \sigma^{\mu\nu} d | M(p) \rangle = \epsilon^{\mu \nu \rho \sigma} \Bigg\{ -\epsilon_\rho^{(\lambda) \,\ast} (p + k)_\sigma T_1(q^2) \cr
&\quad + 2 \frac{\epsilon^*_{(\lambda)} \cdot q}{q^2} p_{\rho} k_{\sigma} \left[T_1(q^2) - T_2(q^2) - \frac{q^2}{m_M^2 - m_V^2} T_3(q^2) \right] \cr
&\quad + \epsilon_\rho^{(\lambda) \,\ast}  q_\sigma \frac{m_M^2 - m_V^2}{q^2} \left(T_1(q^2) - T_2(q^2) \right) \Bigg\},
\end{align}
 with $q=p-k$. The functions that are of main importance for the subject of this paper are $I_{6c}^\pm$ and $I_{6s}^\pm$, as they drive the forward--backward asymmetry of the longitudinal and transverse part of the decay, respectively, cf. Eq.~\eqref{eq:AFBLT2}. We have:
	\begin{align}
 	\label{eq:I6sm}
		&I_{6s}^- = 
		  \frac{\mathcal{N}_{N_L}}{2}\times\cr
		  &\Big[ - \left|4 C^T_{LL} H^T_- K_{+-}^{N_L}  
		+ (C^V_{LL} H^V_{-} - C^V_{RL}  H^V_{+}) 
		K_{++}^{N_L} \right|^2 \cr
		& +\left| 4 C^T_{LL} H^T_+  K_{+-}^{N_L} 
		-( (C^V_{LL} H^V_{+} - C^V_{RL}  H^V_{-})  
		 K_{++}^{N_L} \right|^2\Big]\cr
		 &\qquad + \frac{\mathcal{N}_{N_R}}{2}\times\cr
		&\Big[
		 \left| (C^V_{RR} H^V_{-}- C^V_{LR}H^V_{+}) K_{--}^{N_R}  
		+ 4 C^T_{RR} H^T_- K_{-+}^{N_R}  \right|^2  \cr 
		 & -
		\left| (C^V_{LR} H^V_{-} - C^V_{RR} H^V_{+}) K_{--}^{N_R}
		+ 4 C^T_{RR} H^T_+  K_{-+}^{N_R} \right|^2
		 \Big]\,.\cr
       \end{align}

	\begin{align}
 	\label{eq:I6sp}
		&I_{6s}^+ = 
		  \frac{\mathcal{N}_{N_L}}{2}\times\cr
		  &\Big[ \left|4 C^T_{LL} H^T_- K_{-+}^{N_L}  
		+ (C^V_{LL}  H^V_{-} - C^V_{RL}  H^V_{+}) 
		K_{--}^{N_L} \right|^2 \cr
		& -\left| 4 C^T_{LL} H^T_+  K_{-+}^{N_L} 
		-( C^V_{LL}  H^V_{+} - C^V_{RL}  H^V_{-})  
		 K_{--}^{N_L} \right|^2\Big]\cr
		 &\qquad + \frac{\mathcal{N}_{N_R}}{2}\times\cr
		&\Big[
		 \left| (C^V_{LR} H^V_{-}- C^V_{RR}H^V_{+}) K_{++}^{N_R}  
		+ 4 C^T_{RR} H^T_+ K_{+-}^{N_R}  \right|^2  \cr 
		 & -
		\left| (C^V_{RR} H^V_{-} - C^V_{LR} H^V_{+}) K_{++}^{N_R}
		+ 4 C^T_{RR} H^T_-  K_{+-}^{N_R} \right|^2
		 \Big]\,.\cr
       \end{align}

	\begin{align}
	\label{eq:I6cm}
		&I_{6c}^- = 2 \mathcal{N}_{N_L} 
		\times\cr 
		& \mathrm{Re} \Big\{ \Big[  (- C^V_{LL} + C^V_{RL}) H^V_{t} K_{-+}^{N_L} 
		+ (C^S_{LL} - C^S_{RL}) H^S K_{--}^{N_L}\Big] \cr
		&\qquad  \Big[ (C^V_{LL} - C^V_{RL}) H^V_{0} K_{-+}^{N_L} 
		- 4 C^T_{LL} H^T_{0} K_{--}^{N_L} \Big]\Big\} \cr
		&\qquad+2 \mathcal{N}_{N_R} \times\cr 
		 & \mathrm{Re} \Big\{\Big[ 
		 (C^V_{LR} - C^V_{RR}) H^V_{t} K_{+-}^{N_R} 
		+ (-C^S_{LR} + C^S_{RR}) H_S K_{++}^{N_R} \Big] \cr
		&\Big[  (C^V_{LR} - C^V_{RR}) H^V_{0} K_{+-}^{N_R} 
		+ 4 C^T_{RR} H^T_0  K_{++}^{N_R}  \Big] \Big\}\,,\cr
        \end{align}

	\begin{align}
	\label{eq:I6cp}
		&I_{6c}^+ = 2 \mathcal{N}_{N_L} 
		\times\cr 
		& \mathrm{Re} \Big\{ \Big[  (C^V_{LL} - C^V_{RL}) H^V_{t} K_{+-}^{N_L} 
		+ (-C^S_{LL} + C^S_{RL}) H^S K_{++}^{N_L}\Big] \cr
		&\qquad  \Big[ (C^V_{LL} - C^V_{RL}) H^V_{0} K_{+-}^{N_L} 
		- 4 C^T_{LL} H^T_{0} K_{++}^{N_L} \Big]\Big\} \cr
		&\qquad+2 \mathcal{N}_{N_R} \times\cr 
		 & \mathrm{Re} \Big\{\Big[ 
		 (-C^V_{LR} + C^V_{RR}) H^V_{t} K_{-+}^{N_R} 
		+ (C^S_{LR} - C^S_{RR}) H_S K_{--}^{N_R} \Big] \cr
		&\Big[  (C^V_{LR} - C^V_{RR}) H^V_{0} K_{-+}^{N_R} 
		+ 4 C^T_{RR} H^T_0  K_{--}^{N_R}  \Big] \Big\}\,,\cr
        \end{align}
Note that in the case of vector meson in the final state, $M\to V \ell N$, the normalization function used in the expressions for 
$I_i^\pm$, reads 
\begin{align}
\mc{N}_{N}=\frac{3G_F^2 V_{cb}^2}{4096 \pi^4} \frac{\sqrt{\lambda_{MV}}\sqrt{\lambda_{\ell N}}}{m_M^3 q^2} \mathcal{B}(V\rightarrow P\pi).
 \end{align}
 Finally, for the denominator in Eq.~\eqref{eq:AFBLT} the expressions for $I_{1s}^\pm$, $I_{1c}^\pm$,  $I_{2s}^\pm$, and $I_{2c}^\pm$ are also needed, because the relevant 
decay rates are 
\begin{align}
&\Gamma_L^\pm = \frac{8\pi}{3} \int_{q^2_\mathrm{min}}^{q^2_\mathrm{max}} dq^2\left(I_{1c}^\pm -\frac{1}{3} I_{2c}^\pm\right)\,,\cr
&\Gamma_T^\pm = \frac{16\pi}{3} \int_{q^2_\mathrm{min}}^{q^2_\mathrm{max}} dq^2\left(I_{1s}^\pm -\frac{1}{3} I_{2s}^\pm\right)\,.
\end{align}
For completeness they are listed below.
\begin{equation}
    \begin{aligned}
        I_{1c}^{-}& = \mc{N}_{N_L} \times\\
        &\Bigg[ \Big|(C^S_{LL} - C^S_{RL}) H^S K^{N_L}_{--} + \left( -  C^V_{LL} + C^V_{RL} \right) H^V_t K^{N_L}_{-+}\Big|^2 + \\
        & \frac{1}{2}\Big| (C^V_{RL} -  C^V_{LL}) H^V_0 K^{N_L}_{++} + 4 C^T_{LL} H^T_0 K^{N_L}_{+-} \Big|^2 + \\
        &  \frac{1}{2} \Big| (C^V_{LL} - C^V_{RL} ) H^V_0 K^{N_L}_{-+} - 4 C^T_{LL} H^T_0 K^{N_L}_{--} \Big|^2\Bigg]+\\
         &\quad \mc{N}_{N_R} \times\\
         &\Bigg[ \Big|(C^S_{RR} - C^S_{LR}) H^S K^{N_R}_{++} + (C^V_{LR} - C^V_{RR}) H^V_t K^{N_R}_{+-}\Big|^2 + \\
        & \frac{1}{2}\Big| (C^V_{RR} - C^V_{LR}) H^V_0 K^{N_R}_{--} - 4 C^T_{RR} H^T_0 K^{N_R}_{-+} \Big|^2 + \\
        &  \frac{1}{2} \Big| (C^V_{LR} - C^V_{RR}) H^V_0 K^{N_R}_{+-} + 4 C^T_{RR} H^T_0 K^{N_R}_{++} \Big|^2\Bigg]\,,
    \end{aligned}
\end{equation}
\begin{equation}
    \begin{aligned}
        I_{1c}^{+}& = \mc{N}_{N_L} \times\\
        & \Bigg[ \Big|(C^S_{RL} - C^S_{LL}) H^S K^{N_L}_{++} + (C^V_{LL} - C^V_{RL}) H^V_t K^{N_L}_{+-}\Big|^2 + \\
        & \frac{1}{2}\Big| (C^V_{LL} - C^V_{RL}) H^V_0 K^{N_L}_{--} - 4 C^T_{LL} H^T_0 K^{N_L}_{-+} \Big|^2 + \\
        &  \frac{1}{2} \Big| (C^V_{LL} - C^V_{RL}) H^V_0 K^{N_L}_{+-} - 4 C^T_{LL} H^T_0 K^{N_L}_{++} \Big|^2\Bigg] + \\
        & \quad \mc{N}_{N_R} \times \\
        &\Bigg[ \Big|(-C^S_{RR} + C^S_{LR}) H^S K^{N_R}_{--} + (-C^V_{LR} + C^V_{RR}) H^V_t K^{N_R}_{-+}\Big|^2 + \\
        & \frac{1}{2}\Big| (C^V_{LR} - C^V_{RR}) H^V_0 K^{N_R}_{++} + 4 C^T_{RR} H^T_0 K^{N_R}_{+-} \Big|^2 + \\
        &  \frac{1}{2} \Big| (C^V_{LR} - C^V_{RR}) H^V_0 K^{N_R}_{-+} + 4 C^T_{RR} H^T_0 K^{N_R}_{--} \Big|^2\Bigg]\,,
    \end{aligned}
\end{equation}

\begin{equation}
    \begin{aligned}
        I_{1s}^{-}& = \mc{N}_{N_L}\times\\
        &  \Bigg[ \frac{3}{4}\Big| \left(-C^V_{RL} H^V_{+} +  C^V_{LL} H^V_{-} \right) K^{N_L}_{++} + 4 C^T_{LL} H^T_{-} K^{N_L}_{+-}\Big|^2 + \\
        & \frac{1}{4}\Big|  \left( C^V_{RL} H^V_{-} -  C^V_{LL} H^V_{+} \right) K^{N_L}_{-+} + 4 C^T_{LL} H^T_{+} K^{N_L}_{--} \Big|^2 + \\
        &  \frac{3}{4} \Big|  \left(C^V_{RL} H^V_{-} -  C^V_{LL} H^V_{+} \right) K^{N_L}_{++} + 4 C^T_{LL} H^T_{+} K^{N_L}_{+-} \Big|^2 \\
        & \frac{1}{4} \Big|  \left(C^V_{RL} H^V_{-} -  C^V_{LL} H^V_{+} \right) K^{N_L}_{+-} + 4 C^T_{LL} H^T_{+} K^{N_L}_{++} \Big|^2\Bigg] +\\
        &\quad\mc{N}_{N_R} \times \\
        &\Bigg[ \frac{3}{4}\Big| \left( -C^V_{RR} H^V_{+} + C^V_{LR} H^V_{-} \right) K^{N_R}_{--} + 4 C^T_{RR} H^T_{+} K^{N_R}_{-+}\Big|^2 + \\
        & \frac{1}{4}\Big|  \left(C^V_{RR} H^V_{-} - C^V_{LR} H^V_{+} \right) K^{N_R}_{+-} + 4 C^T_{RR} H^T_{-} K^{N_R}_{++} \Big|^2 + \\
        &  \frac{3}{4} \Big|  \left( C^V_{RR} H^V_{-} - C^V_{LR} H^V_{+} \right) K^{N_R}_{--} + 4 C^T_{RR} H^T_{-} K^{N_R}_{-+} \Big|^2 \\
        & \frac{1}{4} \Big|  \left(C^V_{RR} H^V_{+} - C^V_{LR} H^V_{-} \right) K^{N_R}_{+-} - 4 C^T_{RR} H^T_{+} K^{N_R}_{++} \Big|^2\Bigg]\,,
    \end{aligned}
\end{equation}

\begin{equation}
    \begin{aligned}
        I_{1s}^{+}& = \mc{N}_{N_L} \times\\
        & \Bigg[ \frac{3}{4}\Big| \left( -C^V_{LL} H^V_{+} + C^V_{RL} H^V_{-} \right) K^{N_L}_{--} + 4 C^T_{LL} H^T_{+} K^{N_L}_{-+}\Big|^2 + \\
        & \frac{1}{4}\Big|  \left(C^V_{RL} H^V_{+} -  C^V_{LL} H^V_{-} \right) K^{N_L}_{+-} - 4 C^T_{LL} H^T_{-} K^{N_L}_{++} \Big|^2 + \\
        &  \frac{3}{4} \Big|  \left(  C^V_{LL} H^V_{-} - C^V_{RL} H^V_{+} \right) K^{N_L}_{--} + 4 C^T_{LL} H^T_{-} K^{N_L}_{-+} \Big|^2 \\
        & \frac{1}{4} \Big|  \left(C^V_{RL} H^V_{-} -  C^V_{LL} H^V_{+} \right) K^{N_L}_{+-} + 4 C^T_{LL} H^T_{+} K^{N_L}_{++} \Big|^2\Bigg]+\\
       & \quad \mc{N}_{N_R}\times\\
       & \Bigg[ \frac{3}{4}\Big| \left(-C^V_{LR} H^V_{+} + C^V_{RR} H^V_{-} \right) K^{N_R}_{++} + 4 C^T_{RR} H^T_{-} K^{N_R}_{+-}\Big|^2 + \\
        & \frac{1}{4}\Big|  \left( C^V_{RR} H^V_{+} - C^V_{LR} H^V_{-} \right) K^{N_R}_{-+} - 4 C^T_{RR} H^T_{+} K^{N_R}_{--} \Big|^2 + \\
        &  \frac{3}{4} \Big|  \left(C^V_{LR} H^V_{-} - C^V_{RR} H^V_{+} \right) K^{N_R}_{++} + 4 C^T_{RR} H^T_{+} K^{N_R}_{+-} \Big|^2 \\
        & \frac{1}{4} \Big|  \left( C^V_{RR} H^V_{-} - C^V_{LR} H^V_{+} \right) K^{N_R}_{-+} + 4 C^T_{RR} H^T_{-} K^{N_R}_{--} \Big|^2\Bigg]\,,
    \end{aligned}
\end{equation}

\begin{equation}
    \begin{aligned}
        I_{2c}^{-} &= \mc{N}_{N_L}\times\\\\
         & \Bigg[ -\frac{1}{2} \Big| (C^V_{RL} -  C^V_{LL}) H^V_0 K^{N_L}_{++} + 4 C^T_{LL} H^T_0 K^{N_L}_{+-} \Big|^2 + \\
        & \frac{1}{2}\Big| ( C^V_{LL} - C^V_{RL} ) H^V_0 K^{N_L}_{-+} - 4 C^T_{LL} H^T_0 K^{N_L}_{--} \Big|^2  \Bigg]+\\
        &\quad \mc{N}_{N_R} \times\\
        &\Bigg[ -\frac{1}{2} \Big| (C^V_{RR} - C^V_{LR}) H^V_0 K^{N_R}_{--} - 4 C^T_{RR} H^T_0 K^{N_R}_{-+} \Big|^2 + \\
        & \frac{1}{2}\Big| (C^V_{LR} - C^V_{RR}) H^V_0 K^{N_R}_{+-} + 4 C^T_{RR} H^T_0 K^{N_R}_{++} \Big|^2  \Bigg]\,,
    \end{aligned}
\end{equation}

\begin{equation}
    \begin{aligned}
        I_{2c}^{+} &= \mc{N}_{N_L} \times\\
        & \Bigg[ -\frac{1}{2} \Big| ( C^V_{LL} - C^V_{RL}) H^V_0 K^{N_L}_{--} - 4 C^T_{LL} H^T_0 K^{N_L}_{-+} \Big|^2 + \\
        & \frac{1}{2}\Big| (C^V_{LL} - C^V_{RL}) H^V_0 K^{N_L}_{+-} - 4 C^T_{LL} H^T_0 K^{N_L}_{++} \Big|^2  \Bigg]+\\
        &\quad \mc{N}_{N_R} \times\\
        &\Bigg[ -\frac{1}{2} \Big| (C^V_{LR} - C^V_{RR}) H^V_0 K^{N_R}_{++} + 4 C^T_{RR} H^T_0 K^{N_R}_{+-} \Big|^2 + \\
        & \frac{1}{2}\Big| (C^V_{LR} - C^V_{RR}) H^V_0 K^{N_R}_{-+} + 4 C^T_{RR} H^T_0 K^{N_R}_{--} \Big|^2 \Bigg]\,,
    \end{aligned}
\end{equation}

\begin{equation}
    \begin{aligned}
        I_{2s}^{-}& = \mc{N}_{N_L} \times\\
        &\Bigg[ \frac{1}{4}\Big| \left(-C^V_{RL} H^V_{+} +  C^V_{LL} H^V_{-} \right) K^{N_L}_{++} + 4 C^T_{LL} H^T_{-} K^{N_L}_{+-}\Big|^2 + \\
        & -\frac{1}{4}\Big|  \left( C^V_{RL} H^V_{-} -  C^V_{LL} H^V_{+} \right) K^{N_L}_{-+} + 4 C^T_{LL} H^T_{+} K^{N_L}_{--} \Big|^2 + \\
        &  \frac{1}{4} \Big|  \left(C^V_{RL} H^V_{-} -  C^V_{LL} H^V_{+} \right) K^{N_L}_{++} + 4 C^T_{LL} H^T_{+} K^{N_L}_{+-} \Big|^2 \\
        & -\frac{1}{4} \Big|  \left(C^V_{RL} H^V_{-} -  C^V_{LL} H^V_{+} \right) K^{N_L}_{+-} + 4 C^T_{LL} H^T_{+} K^{N_L}_{++} \Big|^2\Bigg]+\\
        &\quad \mc{N}_{N_R} \times\\
        &\Bigg[ \frac{1}{4}\Big| \left( -C^V_{RR} H^V_{+} + C^V_{LR} H^V_{-} \right) K^{N_R}_{--} + 4 C^T_{RR} H^T_{+} K^{N_R}_{-+}\Big|^2 + \\
        & -\frac{1}{4}\Big|  \left(C^V_{RR} H^V_{-} - C^V_{LR} H^V_{+} \right) K^{N_R}_{+-} + 4 C^T_{RR} H^T_{-} K^{N_R}_{++} \Big|^2 + \\
        &  \frac{1}{4} \Big|  \left( C^V_{RR} H^V_{-} - C^V_{LR} H^V_{+} \right) K^{N_R}_{--} + 4 C^T_{RR} H^T_{-} K^{N_R}_{-+} \Big|^2 \\
        & -\frac{1}{4} \Big|  \left(C^V_{RR} H^V_{+} - C^V_{LR} H^V_{-} \right) K^{N_R}_{+-} - 4 C^T_{RR} H^T_{+} K^{N_R}_{++} \Big|^2\Bigg]\,
    \end{aligned}
\end{equation}

\begin{equation}
    \begin{aligned}
        I_{2s}^{+} &= \mc{N}_{N_L} \times\\
        &\Bigg[ \frac{1}{4}\Big| \left( -C^V_{LL} H^V_{+} + C^V_{RL} H^V_{-} \right) K^{N_L}_{--} + 4 C^T_{LL} H^T_{+} K^{N_L}_{-+}\Big|^2 + \\
        & -\frac{1}{4}\Big|  \left(C^V_{RL} H^V_{+} -  C^V_{LL} H^V_{-} \right) K^{N_L}_{+-} - 4 C^T_{LL} H^T_{-} K^{N_L}_{++} \Big|^2 + \\
        &  \frac{1}{4} \Big|  \left(  C^V_{LL} H^V_{-} - C^V_{RL} H^V_{+} \right) K^{N_L}_{--} + 4 C^T_{LL} H^T_{-} K^{N_L}_{-+} \Big|^2 \\
        & -\frac{1}{4} \Big|  \left(C^V_{RL} H^V_{-} -  C^V_{LL} H^V_{+} \right) K^{N_L}_{+-} + 4 C^T_{LL} H^T_{+} K^{N_L}_{++} \Big|^2\Bigg]+\\
        &\quad \mc{N}_{N_R} \times \Bigg[ \frac{1}{4}\Big| \left(-C^V_{LR} H^V_{+} + C^V_{RR} H^V_{-} \right) K^{N_R}_{++} + 4 C^T_{RR} H^T_{-} K^{N_R}_{+-}\Big|^2 + \\
        & -\frac{1}{4}\Big|  \left( C^V_{RR} H^V_{+} - C^V_{LR} H^V_{-} \right) K^{N_R}_{-+} - 4 C^T_{RR} H^T_{+} K^{N_R}_{--} \Big|^2 + \\
        &  \frac{1}{4} \Big|  \left(C^V_{LR} H^V_{-} - C^V_{RR} H^V_{+} \right) K^{N_R}_{++} + 4 C^T_{RR} H^T_{+} K^{N_R}_{+-} \Big|^2 \\
        & -\frac{1}{4} \Big|  \left( C^V_{RR} H^V_{-} - C^V_{LR} H^V_{+} \right) K^{N_R}_{-+} + 4 C^T_{RR} H^T_{-} K^{N_R}_{--} \Big|^2\Bigg]\,,
    \end{aligned}
\end{equation}



\end{document}